\documentclass[journal]{IEEEtran}
\usepackage[utf8]{inputenc}
\usepackage{amsmath,amssymb,amsfonts}
\usepackage{algorithm} 
\usepackage{epstopdf}
\usepackage{subcaption} 
\usepackage{caption}    
\usepackage{textcomp}
\usepackage{bm}
\usepackage{amsmath,amsfonts}
\usepackage{algorithmic}
\usepackage{array}
\usepackage{textcomp}
\usepackage{stfloats}
\usepackage{url}
\usepackage{verbatim}
\usepackage{graphicx}
\usepackage{booktabs}
\usepackage{balance}
\usepackage{multirow}
\usepackage{diagbox}
\usepackage{float}
\usepackage{xcolor}
\def\BibTeX{{\rm B\kern-.05em{\sc i\kern-.025em b}\kern-.08em
		T\kern-.1667em\lower.7ex\hbox{E}\kern-.125emX}}
\allowdisplaybreaks

\begin{document}
	\title{\huge Energy-Aware Task Offloading for Rotatable STAR-RIS-Enhanced Mobile Edge Computing Systems}
	\author{Dongdong Yang,~Bin Li,~\IEEEmembership{Member,~IEEE},~and Dusit Niyato,~\IEEEmembership{Fellow,~IEEE}
		
		\thanks{Dongdong Yang and Bin Li are with the School of Computer Science, Nanjing University of Information Science and Technology, Nanjing 210044, China, and also with the Huzhou Key Laboratory of Urban Multidimensional Perception and Intelligent Computing, Huzhou 313000, China (e-mail:  202312200024@nuist.edu.cn; bin.li@nuist.edu.cn).}
		
		\thanks{Dusit Niyato is with the College of Computing and Data Science, Nanyang Technological University, Singapore (e-mail: dniyato@ntu.edu.sg).}
		}

	\setlength{\parskip}{0pt} 
	\maketitle
	
	\begin{abstract}
Simultaneously transmitting and reflecting reconfigurable intelligent surface (STAR-RIS) can expand the coverage of mobile edge computing (MEC) services by reflecting and transmitting signals simultaneously, enabling full-space coverage. The orientation of the STAR-RIS plays a crucial role in optimizing the gain of received and transmitted signals, and a rotatable STAR-RIS offers potential enhancement for MEC systems. This paper investigates a rotatable STAR-RIS-assisted MEC system, operated under three protocols, namely energy splitting, mode switching, and time switching. The goal is to minimize energy consumption for multiple moving user devices through the joint optimization of STAR-RIS configurations, orientation, computation resource allocation, transmission power, and task offloading strategies. Considering the mobility of user devices, we model the original optimization problem as a sequential decision-making process across multiple time slots. The high-dimensional, highly coupled, and nonlinear nature makes it a challenging non-convex decision-making problem for traditional optimization algorithms. Therefore, a deep reinforcement learning (DRL) approach is employed, specifically utilizing soft actor-critic algorithm to train the DRL model. Simulation results demonstrate that the proposed algorithm outperforms the benchmarks in both convergence speed and energy efficiency, while reducing energy consumption by up to 52.7\% compared to the fixed STAR-RIS scheme. Among three operating protocols, the energy splitting yields the best performance.
	
	\end{abstract}
	
	\section{Introduction}
	With the exponential growth of data generated by numerous user devices (UDs) and the increasing demands for computation-intensive and latency-sensitive applications, traditional cloud computing systems have been proven insufficient for the evolving Internet of Things (IoT) era. Therefore, mobile edge computing (MEC) has emerged as a promising paradigm that enables UDs to offload tasks to nearby edge nodes, significantly reducing reliance on distant data centers \cite{10207705}. For instance, in smart manufacturing, MEC servers, strategically deployed near the production lines, can process data in real-time, enabling immediate quality control and equipment monitoring, thereby optimizing manufacturing processes \cite{ref3}. Similarly, in augmented reality, MEC  can enhance the user experience by providing high-speed computational support proximate to UDs \cite{ref4}. However, MEC systems still face challenges like non-line-of-sight (NLoS) radio propagation, leading to potential task offloading failures or excessive delays.
	
	Fortunately, reconfigurable intelligent surface (RIS), valued for its passivity, cost-effectiveness, and reconfigurability, has been recognized as an essential technology for 6G \cite{9615187}. Different from traditional passive adaptation, where terminal devices adjust to the wireless environment passively, RIS actively configures the communication environment to enhance signal quality \cite{ref6}. Specifically, each RIS consists of numerical low-cost passive reflecting elements, digitally regulated to induce the amplitude and phase shifts of incident signals, achieving passive beamforming to enhance or suppress signals in specific direction \cite{ref7}. When direct links between UDs and base station (BS) are blocked, RIS can establish virtual line-of-sight (LoS) links, effectively mitigating NLoS issues and improving communication quality. Moreover, with the improvement brought by RIS, the transmission power consumption of UDs is significantly reduced, further enhancing energy efficiency. By leveraging these advantages, RIS can strengthen offloading links and provide substantial benefits to MEC systems \cite{ref8}.
	
	However, conventional RIS is limited to either reflecting or transmitting incident signals, resulting in half-space coverage where both the BS and UDs must be positioned either on the same side (reflecting-only RIS) or the opposite sides (transmitting-only RIS) \cite{ref9}. This spatial limitation greatly restricts the flexibility of RIS deployment and is often impractical. To address this, simultaneously transmitting and reflecting RIS (STAR-RIS) was introduced \cite{ref11}, dividing incident signals into two parts: one reflected back into the same half-space and the other transmitted to the opposite half-space. This full-space coverage capability allows STAR-RIS-assisted MEC networks to overcome spatial limitations, providing greater coverage and supporting continuous, high-quality services \cite{ref12}.

Recent studies have shown that the actual gain of the STAR-RIS is influenced by its orientation, and appropriately adjusting its orientation can optimize the incident and reflected angles of the signal, thereby maximizing the performance gain of the STAR-RIS. However, the orientation optimization remains challenging. On one hand, both the receiving and transmitting gains with respect to wave incidence and transmission are sensitive to its orientation. On the other hand, the phase shifts of the STAR-RIS are tightly coupled with its orientation \cite{ref13,ref14}. To address this, we propose a rotatable STAR-RIS scheme to further enhance the flexibility and efficiency of MEC networks. By introducing rotation, the STAR-RIS can dynamically adjust its orientation to optimize the communication environment, thereby maximizing offloading efficiency and improving overall network performance \cite{ref15}. However, several challenges must be overcome to harmoniously integrate the rotatable STAR-RIS into MEC systems \cite{ref16}. For example, although RIS configurations have been intensively studied, existing algorithms are not suitable of the optimization of STAR-RIS configurations since they must be designed based on different operating protocols, namely energy splitting (ES), mode switching (MS), and time switching (TS) \cite{ref11}. Meanwhile, STAR-RIS introduces more adjustable parameters beyond phase shift, i.e., the amplitude coefficients for ES/MS protocol, and time allocation for TS protocol, all of which are highly coupled with STAR-RIS orientation and computation resource allocation for MEC \cite{ref17}. In addition, although existing studies have explored the orientation optimization of RIS/STAR-RIS, most of them focus on static- or single-UD environments \cite{ref18,ref19}, which may not fully capture the complexities arising from multiple moving UDs in MEC systems. Therefore, this paper proposes a rotatable STAR-RIS scheme to account for dynamic environments and enhance MEC performance. The primary contributions of this paper are outlined as follows.
	  
	\begin{itemize}
		\item We propose a novel rotatable STAR-RIS-assisted MEC system with multiple moving UDs, where the STAR-RIS can dynamically adjust its orientation to accommodate UDs mobility. Considering the characteristics of STAR-RIS, the energy consumption minimization problem is formulated for three operating protocols, jointly optimizing STAR-RIS configurations, orientation, computation resource allocation, transmission power, and task offloading strategies.
		\item To account for practical scenarios, we incorporate discrete phase shift, STAR-RIS elements grouping, and UDs mobility into our model. Thus, the energy consumption minimization problem is formulated as a time-vary problem. Considering this problem is difficult to handle using traditional algorithms, we propose a sample-efficient deep reinforcement learning (DRL)  algorithm based on soft actor-critic (SAC) to adapt to the dynamic environment and achieve near-optimal solutions. 
		\item Simulation results validate that the proposed approach outperforms representative DRL algorithms in both convergence and performance. Additionally, the proposed rotatable STAR-RIS scheme significantly reduces energy consumption compared to the fixed scheme.
	\end{itemize}
	
	The rest of the paper is structured as follows. Related works are discussed in Section II. In Section III, the system model and formulated energy consumption minimization problems are presented. In Section IV, we detail the proposed solution algorithms. In Section V, performance evaluation is conducted to demonstrate the validation of our proposed system. Finally, Section VI concludes this paper.
	
	\textit{Notation:} Throughout this paper, non-boldface, boldface lowercase, and boldface uppercase letters denote scalar, vector, and matrix, respectively; $\left| {\cdot} \right|$, ${\left\| {\cdot} \right\|_2}$, and ${\left( {\cdot} \right)^{\rm{H}}}$ denote absolute value, $\ell_2$-norm, respectively; $\mathbb{E}\left( {\cdot} \right)$, $\text{diag}\left({\mathbf a} \right)$, ${\cal CN}(0,\sigma^2)$ represent expectation, a diagonal matrix with vector $\mathbf a$ on the main diagonal, and complex Gaussian distribution with zero mean value and variance $\sigma^2$, respectively.

	\section{Related Works}
	\subsubsection{MEC}
	In recent years, MEC technologies have garnered significant attention from academic researchers and been applied in various domains \cite{8493155}. By provisioning extensive computation resource at edge, MEC facilitates low-latency wireless access, thereby enabling lightweight terminals with limited processing capability to efficiently handle computing-intensive tasks. Furthermore, researchers have conducted in-depth optimizations of MEC systems with diverse objectives. For example, \cite{ref25} proposed a wireless powered MEC system featuring binary computation offloading, utilizing block coordinate descent (BCD) and alternating direction method of multipliers to maximize computation rate by optimizing the selection of computing mode and allocation of transmission time. In scenarios where UDs are highly sensitive to latency, \cite{ref23} focused on weighted total latency minimization and proposed an innovative approach jointly designing communication, computation, and caching configurations in MEC systems. To improve task execution efficiency, \cite{ref24} investigated the energy efficiency maximization problem for task offloading in a two-tier MEC network. Nevertheless, the aforementioned studies mainly concentrate on task offloading for fixed UDs, which may limit their applicability in practical scenarios. To address this, \cite{ref27} introduced a mobility-aware MEC network and employed reinforcement learning (RL) to jointly optimize task offloading and migration schemes aiming to maximize system revenue. Additionally, to further extend MEC network coverage and improve deployment flexibility, \cite{ref26} incorporated unmanned aerial vehicles (UAVs) into MEC systems to maximize computation rate.
	\subsubsection{RIS and STAR-RIS}
	Thanks to its advantageous features, RIS has garnered substantial attention from both industry and academia. For example, \cite{latencyRIS} jointly designed phase shifts, offloading strategies, and transmission power based on BCD and Lagrange multiplier method to minimize weighted total latency. To further ensure task offloading security, \cite{secureRIS} proposed a RIS-assisted MEC system framework, and investigated the problem of maximizing the minimum computation efficiency, optimizing local computation frequencies, IoT devices' transmission power and RIS phase shifts to meet secure computation rate demands. For moving UDs, \cite{bin} aimed to minimize energy consumption through jointly designing RIS phase shifts and UDs' transmission power.
	
	Despite these appealing features of RIS, the spatial restriction of transmitters and receivers imposes challenges in practical implementation, prompting the development of STAR-RIS \cite{ref11,ref12}. Currently, the study of STAR-RIS is still in its early stages. In \cite{ref31}, Xu \textit{et al}. proposed the STAR-RIS concept, introducing standard hardware and channel models covering both near-field and far-field scenarios. Then, the operating protocols of STAR-RIS were further investigated in \cite{ref12}, analyzing the constraints, merit, and demerit of each protocol. Furthermore, in \cite{ref32},  STAR-RIS was introduced into MEC systems, where the STAR-RIS was operated under the TS protocol, with the objective of minimizing the total energy consumption of UDs, involving the joint optimization of STAR-RIS transmission/reflection time and coefficients. Similarly, in \cite{ref33}, Liu \textit{et al}. discussed the STAR-RIS-assisted MEC systems under the MS protocol and formulated a computation rate maximization problem via jointly designing receive beamforming, STAR-RIS configurations, and energy partition strategies for offloading and local computing. Simulations demonstrated that STAR-RIS outperforms conventional RIS, confirming its superiority.
	\subsubsection{Orientation optimization for RIS/STAR-RIS}
	Numerous studies have demonstrated the significant influence of RIS/STAR-RIS orientation on overall system. In \cite{ref13}, Cheng \textit{et al}. conducted an in-depth investigation into the impact of RIS orientation and deployment on wireless communications, introducing RIS rotation as an additional degree of freedom in RIS-assisted systems. Simulation results demonstrated the effectiveness of rotation, and in certain scenarios, rotating the RIS yields greater gains than moving it. Furthermore, in \cite{ref14}, Wang \textit{et al}. explored the rotation of STAR-RIS and utilized deep learning to optimize STAR-RIS orientation in various scenarios, achieving full-space coverage while maximizing STAR-RIS gain. Simulations have shown that the rotatable STAR-RIS is superior to the fixed, verifying the advantage of real-time orientation optimization.
	
	Many researchers have attempted to integrate RIS into UAVs to expand its coverage, allowing for some adjustment of RIS orientation as the UAV moves \cite{10243608}. Nevertheless, due to the multiplicative fading and the considerable distance between UAVs and transmitters/receivers, the actual performance of UAV-RIS system remains limited \cite{aerial_RIS,active_RIS}. Additionally, the high energy consumption associated with UAV flight further constrains the practical effectiveness. Moreover, some researchers have combined fluid antennas with RIS, proposing the concept of movable RIS to eliminate phase distribution offset by the flexible adjustment of each RIS element position \cite{movable_RIS}. Building on these findings, we are inspired to focus on enhancing RIS gain by RIS rotation rather than spatial repositioning. However, to the best of our knowledge, most works have overlooked the potential of rotatable STAR-RIS in enhancing channels within MEC systems, and thereby improving task offloading performance. 
	\section{System Model and Problem Formulation}\label{s:sys}
	In this section, we first present three operating protocols for STAR-RIS in MEC systems. Then, based on these protocols, we introduce the system model of a rotatable STAR-RIS-assisted MEC system. Finally, we formulate the joint optimization problem involving STAR-RIS orientation, configurations, transmission power, and offloading strategies for multiple moving UDs. Table I presents all the important symbols.
	
	\begin{table}[htbp]
		\renewcommand{\arraystretch}{1.2} 
		\centering
		\caption{\scriptsize {LIST OF VARIABLES}}
		\begin{tabular}{@{}c@{\hspace{0.02\textwidth}}p{0.40\textwidth}@{}}
			\toprule
			Variable & Description \\
			\midrule
			$K$ & $\text{The number of UDs}$ \\
			$N$ & $\text{The number of STAR-RIS elements}$ \\
			$\bar N$ & $\text{The number of sub-surfaces}$ \\
			$T$ & $\text{The duration of task cycle}$\\
			$Q$ & $\text{The number of time slots}$\\
			$D_k$ & $\text{The task size of UD }k$\\
			$C_k$ & $\text{The required CPU cycles of task }D_k$\\
			$\tau$ & $\text{The length of each time slot}$\\
			$h_{k,\rm{R}}^n$ & $\text{The channel between UD $k$ and the $n$-th STAR-RIS element}$\\
			$\bar{h}_{k,\rm{R}}^n/\tilde{h}_{k,\rm{R}}^n$ & $\text{The LoS/NLoS component of $h_{k,\rm{R}}^n$}$\\
			$v_{\rm{R,B}}^n$ & $\text{The channel between the $n$-th element and BS}$\\
			$\bar{v}_{\rm{R,B}}^n/\tilde{v}_{\rm{R,B}}^n$ & $\text{The LoS/NLoS component of $v_{\rm{R,B}}^n$}$\\
			$K_1/K_1$ & $\text{The Rician factor of channels between UDs/BS and STAR-RIS}$\\
			$\rho_0$ & $\text{The pass-loss factor at a reference distance of 1 meter}$\\
			$d_{k,{\rm R}}^n/d_{{\rm R,B}}^n$ & $\text{The distance between UD $k$/BS and the $n$-th element}$\\
			$\mathbf{\Phi_{\rm t}/\Phi_{\rm r}}$ & $\text{The transmission/reflection coefficient matrix of the STAR-RIS}$\\
			$\beta_n^{\rm t}/\beta_n^{\rm r}$ & $\text{The transmission/reflection amplitude coefficient}$\\
			$\phi_n^{\rm t}/\phi_n^{\rm r}$ & $\text{The transmission/reflection phase shift of the $n$-th element}$\\
			$\lambda_{\rm t}/\lambda_{\rm r}$ & $\text{The operating mode of STAR-RIS under the TS protocol}$\\
			$u_k$ & $\text{The parameter indicating whether UD $k$ is within RA or TA}$\\
			$F\left( {\upsilon,\vartheta} \right)$ & $\text{The normalized power radiation pattern of the STAR-RIS}$\\
			$\delta$ & $\text{The rotation angle}$\\
			$\theta_k/\theta_{\rm B}$ & $\text{The angle from UD $k$/BS to the STAR-RIS projected on XOY}$\\
			$\varphi_k/\varphi_{\rm B}$ & $\text{The elevation angle between UD $k$/BS and the STAR-RIS}$\\
			$G_k/G_{\rm B}$ & $\text{The receiving/transmitting gain}$\\
			$D_{\rm m}$ & $\text{The maximum directivity of the STAR-RIS}$\\
			\bottomrule
		\end{tabular}
		\label{tab:notation}
	\end{table}

	\renewcommand{\arraystretch}{1.5} 
	
	\begin{table*}[htbp]
		\centering
		\caption{\scriptsize {OPTIMIZATION VARIABLES AND CONSTRAINTS FOR THREE  PROTOCOLS.}}
		\vspace{-0.1cm}
		\resizebox{0.9\textwidth}{!}{
		\begin{tabular}{|c|c|c|c|}
			\hline
			\textbf{Optimization Variables} & \textbf{Energy splitting} & \textbf{Mode switching} & \textbf{Time switching} \\ 
			\hline
			Phase shift coefficients & \multicolumn{3}{c|}{\(\phi^{\rm t}_n, \phi^{\rm r}_n \in [0, 2\pi)\)} \\ 
			\hline
			Amplitude coefficients 
			& \(\beta^{\rm t}_n, \beta^{\rm r}_n \in [0,1]; \beta^{\rm t}_n + \beta^{\rm t}_n = 1\) 
			& \(\beta^{\rm t}_n, \beta^{\rm r}_n \in \{0,1\}; \beta^{\rm t}_n + \beta^{\rm r}_n = 1\) 
			& \diagbox{}{} \\ 
			\hline
			Time allocation & \diagbox{}{} & \diagbox{}{} & \(\lambda^{\rm t}, \lambda^{\rm r} \in [0,1]; \lambda^{\rm t} + \lambda^{\rm r} = 1\) \\ 
			\hline
		\end{tabular}
	}
	\end{table*}
	
	\subsection{Three practical operating protocols for STAR-RIS}
	Different from conventional RIS with non-magnetic elements, STAR-RIS introduces equivalent surface electric and magnetic currents, enabling transmitted and reflected signals to be interpreted as waves emitted from the time-varying surface equivalent electric and magnetic currents. As shown in Fig. 1, by designing the amplitude coefficients for reflection and transmission, a STAR-RIS element is operated in transmission mode (T mode), reflection mode (R mode), or simultaneous transmission and reflection mode (T$\&$R mode). According to this, we adopt three protocols for STAR-RIS in MEC systems, namely ES, MS, and TS.
	\begin{figure}[t]
		\centering
		\includegraphics[width=0.5\textwidth]{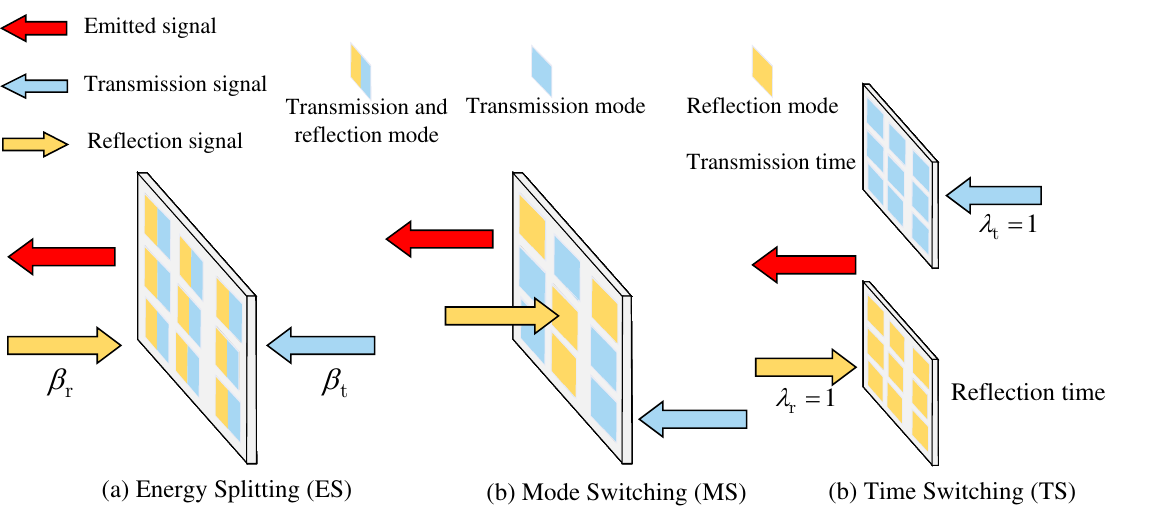}
		\captionsetup{justification=raggedright, singlelinecheck=false} 
		\caption{Three operating protocols for STAR-RIS.}
		\label{fig:star-ris}
	\end{figure}
	\subsubsection{ES}
	In the ES protocol, all elements are operated in T$\&$R mode during task offloading. The amplitude coefficients for reflection and transmission of the $n$-th element, denoted as $\beta_n^{\rm r}$ and $\beta_n^{\rm t}$, satisfy $\beta_n^{\rm t}, \beta_n^{\rm r} \in [0,1]$ and $\beta_n^{\rm t}+\beta_n^{\rm r}=1$. Similarly, the phase shift coefficients for reflection and transmission are $\phi_n^{\rm r}$ and $\phi_n^{\rm t}$, with $\phi_n^{\rm r},\phi_n^{\rm t} \in [0,2\pi)$. Therefore, the STAR-RIS reflection and transmission coefficient matrices are given by ${{\bf{\Phi }}_{\rm{r}}} = {\rm{diag}}\left( {\sqrt {\beta _1^{\rm{r}}} {e^{j\phi _1^{\rm{r}}}}, \ldots ,\sqrt {\beta _n^{\rm{r}}} {e^{j\phi _n^{\rm{r}}}}, \ldots ,\sqrt {\beta _N^{\rm{r}}} {e^{j\phi _N^{\rm{r}}}}} \right)$ and ${{\bf{\Phi }}_{\rm{t}}} = {\rm{diag}}\left( {\sqrt {\beta _1^{\rm{t}}} {e^{j\phi _1^{\rm{t}}}}, \ldots ,\sqrt {\beta _n^{\rm{t}}} {e^{j\phi _n^{\rm{t}}}}, \ldots ,\sqrt {\beta _N^{\rm{t}}} {e^{j\phi _N^{\rm{t}}}}} \right)$, respectively.

	\subsubsection{MS}
	In the MS protocol, each element is operated in either T or R mode, with binary constraints on amplitude coefficients, i.e., $\beta _n^{\rm{t}},\beta _n^{\rm{r}} \in \{ 0,1\}$ and $\beta_n^{\rm t}+\beta_n^{\rm r}=1$. The STAR-RIS coefficient matrices operated under MS protocol are identical to those under the ES protocol.
	
	\subsubsection{TS}
	In the TS protocol, all elements are operated in a uniform mode (T or R mode) within a given time slot. Let $\lambda_{\rm t}$ and $\lambda_{\rm r}$ denote the operating mode, where $\lambda_{\rm t}+\lambda_{\rm r}=1$ and $\lambda_{\rm t},\lambda_{\rm r}\in \{ 0,1\}$. When $\lambda_{\rm t}=1$, the STAR-RIS is operated in T mode, and in R mode when $\lambda_{\rm r}=1$. Therefore, the STAR-RIS coefficient matrices operated under the TS protocol are given by ${{\bf{\Phi }}_{\rm{r}}} = {\lambda _{\rm{r}}}{\rm{diag}}\left( {{e^{j\phi _1^{\rm{r}}}}, \ldots ,{e^{j\phi _n^{\rm{r}}}}, \ldots ,{e^{j\phi _N^{\rm{r}}}}} \right)$ and ${{\bf{\Phi }}_{\rm{t}}} = {\lambda _{\rm{t}}}{\rm{diag}}\left( {{e^{j\phi _1^{\rm{t}}}}, \ldots ,{e^{j\phi _n^{\rm{t}}}}, \ldots ,{e^{j\phi _N^{\rm{t}}}}} \right)$, respectively. However, UDs located in the transmission area (TA) or reflection area (RA) struggle to offload tasks when the STAR-RIS is operated in the opposite mode, causing increased energy consumption.

	TABLE II summarizes the optimization variables and constraints for all three operating protocols.
	
	\subsection{System model}
\begin{figure}[t]
		\centering
		\includegraphics[width=0.8\columnwidth,height=5.2cm]{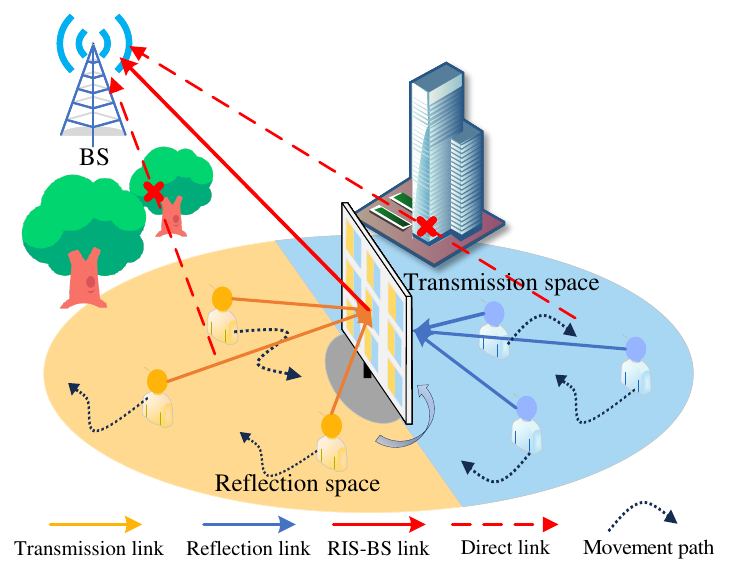}
		\caption{The system model of proposed rotatable STAR-RIS-assisted MEC system.}
		\label{fig:sys-model}
	\end{figure}

	In this paper, we consider UDs operating within a task cycle of duration $T$, divided into $Q$ intervals, each with length $\tau = T/Q$. The set of time slots is represented by $\mathcal{Q}=\{ 1,\ldots,q,\ldots,Q\}$. It is assumed that UDs remain stationary within each time slot and only change their position between slots. As shown in Fig. \ref{fig:sys-model}, we consider a rotatable STAR-RIS-assisted MEC system, where a single-antenna BS equipped with an MEC server supports task offloading for $K$ single-antenna moving UDs. Given the high likelihood of blockage in direct links between UDs and the BS, an $N$-element STAR-RIS is strategically deployed to simultaneously establish reflection or transmission links. Moreover, the STAR-RIS can flexibly adjust its orientation based on the relative positions of the BS and UDs. The sets of UDs and STAR-RIS elements are represented by $\mathcal{K}=\{ 1,...,k,...,K\}$ and $\mathcal{N}=\{ 1,...,n,...,N\}$, respectively. 
	
	To uniquely define the angle of UD $k$ (BS) relative to the STAR-RIS, one half-plane of the STAR-RIS is selected as the 0-degree reference. Consequently, the angle of UD $k$ (BS) relative to the STAR-RIS, projected onto the XOY plane and denoted by $\theta_{k(\rm B)}$, lies within the range $[0,2\pi)$. As the STAR-RIS rotates, these angles are calculated as follows
	\begin{equation}
		{\theta _{k({\rm{B}})}} = \left( {\theta _{k({\rm{B}})}^0 - \delta [q] + 2\pi } \right)\bmod 2\pi,
	\end{equation}
	where $\delta [q]$ denotes the rotation angle, and $\theta _{k({\rm{B}})}^0$ represents the angle between UD $k$ (BS) and the STAR-RIS in its initial orientation. Furthermore, as the STAR-RIS rotates and UDs move, some UDs may transition between the RA and TA as shown in Fig. 3. To indicate whether UD $k$ is within the RA or TA, $u_k$ is introduced and given by
	\begin{equation}
		{u_k}[q] = \left\{ {\begin{array}{*{20}{c}}
				{1,}&{{\theta_k}[q] \in [0,\pi ],}\\
				{0,}&{\text{otherwise},\quad\;}
		\end{array}} \right.
	\end{equation}
	where, when ${u_k}[q]=1$, UD $k$ is in the RA, and in the TA when ${u_k}[q]=0$. For successful task offloading, the BS must remain within the RA, imposing the following constraint on the rotation angle $\delta[q]$
	\begin{equation}
		\delta [q] \in [\theta _{\rm{B}}^0 - \pi ,\theta _{\rm{B}}^0].
	\end{equation}
	
	For simplicity, we assume that UDs move at a constant height $H$. The location of UD $k$ at time slot $q$ is defined as ${\mathbf{\textit{l}}}_{k}[q] = [{x_k}[q], {y_k}[q], H]$, and  updated using the Gauss-Markov random model \cite{bin}, given by
	\begin{equation}
		{x_k}[q + 1] = {x_k}[q] + {v_k}[q]\cos ({\varphi _k}[q])\tau,
	\end{equation}
	\begin{equation}
		{y_k}[q + 1] = {y_k}[q] + {v_k}[q]\sin ({\varphi _k}[q])\tau.
	\end{equation}
	The coordinates of the BS and STAR-RIS are represented as $\bm{\textit{l}}_{\rm B}[q] = [x_{\rm B}, y_{\rm B}, H_{\rm B}]$ and $\bm{\textit{l}}_{\rm R}[q] = [x_{\rm R}, y_{\rm R}, H_{\rm R}]$, respectively. Thus, the elevation angles between the STAR-RIS and UD $k$, as well as between the STAR-RIS and BS are calculated by
	\begin{equation}
		{\varphi _k}[q] = \arcsin \left( {\frac{{\left| {{H_{\rm{R}}} - H} \right|}}{{{{\left\| {{{\mathbf{\textit{l}}}_{\rm{R}}} - {{\mathbf{\textit{l}}}_k}[q]} \right\|}_2}}}} \right),
	\end{equation}
	\begin{equation}
		{\varphi _{\rm B}} = \arcsin \left( {\frac{{\left| {{H_{\rm{B}}} - H_{\rm R}} \right|}}{{{{\left\| {{{\mathbf{\textit{l}}}_{\rm{B}}} - {{\mathbf{\textit{l}}}_{\rm R}}} \right\|}_2}}}} \right).
	\end{equation}

	\begin{figure}[t]
		\centering
		\includegraphics[width=1\columnwidth]{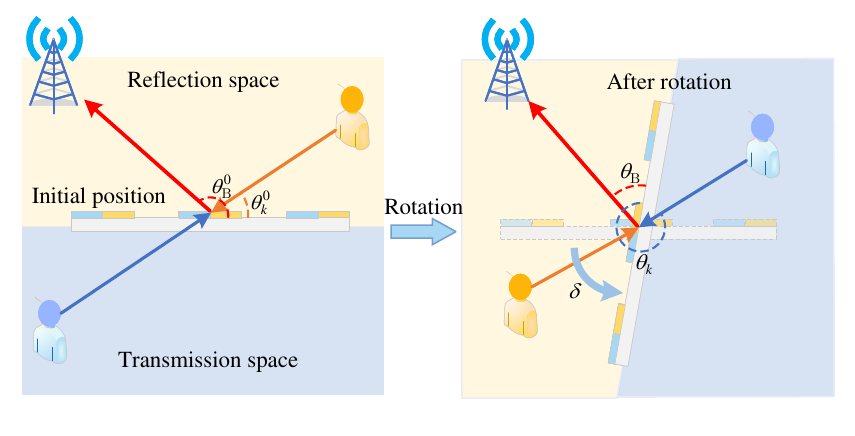}
		\caption{Corresponding angle variations and UDs transition between the RA and TA as the STAR-RIS rotates.}
		\label{fig:rotation}
	\end{figure}
	
	\subsubsection{Communication Model}
	During time slot $q$, the channels between UD $k$ and the STAR-RIS, as well as between the STAR-RIS and BS, are denoted by ${\bm{h}_{k,{\rm{R}}}}[q]$ and $\bm{v}_{\rm{R},\rm{B}}[q]$, respectively. These channels are determined by the individual channels between UD $k$ and each STAR-RIS element $h_{k,\rm{R}}^n[q]$, as well as between each element and the BS $v_{\rm{R},\rm{B}}^n[q]$, both modeled as the Rician distribution \cite{ref16}:
	\begin{equation}
	\scalebox{0.92}{$
		\begin{aligned}
			h_{k,\rm{R}}^n[q] &= \sqrt{\frac{\rho_0}{{d_{k,\rm{R}}^n[q]}^{\alpha_1}}} 
			\left( 
			\sqrt{\frac{K_1}{1 + K_1}} \bar{h}_{k,\rm{R}}^n[q] + 
			\sqrt{\frac{1}{1 + K_1}} \tilde{h}_{k,\rm{R}}^n[q] 
			\right),
		\end{aligned}
		$}
	\end{equation}
	\begin{equation}
	\scalebox{0.92}{$
		\begin{aligned}
			v_{\rm{R},\rm{B}}^n[q] &= \sqrt{\frac{\rho_0}{{d_{\rm{R},\rm{B}}^n[q]}^{\alpha_2}}} 
			\left( 
			\sqrt{\frac{K_2}{1 + K_2}} \bar{v}_{\rm{R},\rm{B}}^n[q] + 
			\sqrt{\frac{1}{1 + K_2}} \tilde{v}_{\rm{R},\rm{B}}^n[q] 
			\right),
		\end{aligned}
		$}
\end{equation}
where ${\rho_0}$ signifies the path-loss factor at a reference distance of 1 meter,  $d^n_{k,\rm{R}}[q]$ ($d^n_{\rm{R},\rm{B}}[q]$) denotes the distance between UD $k$ (BS) and the STAR-RIS. The pass-loss exponents are $\alpha_1$ and $\alpha_2$, while $K_1$ and $K_2$ represent the Rician factors. The LoS components are given by ${{\bar h}_{k,\rm{R}}^n}[q]=\exp \left(-j\frac{2\pi}{\lambda} d_{k,\rm{R}}^n[q]\right)$ and ${{\bar v}_{\rm{R},\rm{B}}^n}[q]=\exp \left(-j\frac{2\pi}{\lambda} d_{\rm{R},\rm{B}}^n[q]\right)$, while the NLoS components follow $\tilde h_{k,{\rm{R}}}^n[q] \sim {\cal CN}(0,1)$ and $\tilde v_{{\rm{R}},{\rm{B}}}^n[q] \sim {\cal CN}(0,1)$. Since the LoS paths between UDs and the BS are obstructed, the channels are modeled as Rayleigh fading channels.
	
Due to practical hardware limitations, the discrete phase shift must be considered \cite{ref21}. Therefore, the $b$-bit discrete phase shift is characterized by the set
\begin{equation}
	\scalebox{0.96}{$
		{\cal F} = \left\{ {\phi _n^m[q]|\phi _n^m[q] \in \left\{ {0,{2^{1 - b}}\pi ,...,(2 - {2^{1 - b}})\pi } \right\}} \right\},m \in \left\{ {{\rm{r}},{\rm{t}}} \right\}.
	$}
\end{equation}
Moreover, the numerical STAR-RIS elements may cause significant complexity in acquiring channel state information and designing STAR-RIS configurations. To reduce this complexity, following the approach in \cite{ref21}, the $N$ elements are grouped into $\bar N$ sub-surfaces, represented by the set $\mathcal{\bar N} = \{ 1, \ldots,\bar n,\ldots,\bar N\}$, where each sub-surface consists of $N/\bar N$  contiguous elements sharing same configurations.
	
Since UDs and the BS are positioned in the far-field region of the STAR-RIS, the distances and angles between UDs (BS) and each element are considered roughly uniform. Thus, the STAR-RIS gain for UD $k$ is given by \cite{ref13}
	\begin{equation}
		\begin{array}{l}
			{{\boldsymbol{\xi}}_k} = {G_k}[q]{G_{\rm{B}}}[q]{{\bf{\Phi }}_m}[q]\\
			\quad {\kern 1pt}  \buildrel \Delta \over = D_{\rm{m}}^2F\left( {\upsilon _{k,{\rm{R}}}^{{\rm{AOD}}}[q],\vartheta _{k,{\rm{R}}}^{{\rm{AOD}}}[q]} \right)F\left( {\upsilon _{{\rm{R,B}}}^{{\rm{AOA}}},\vartheta _{{\rm{R,B}}}^{{\rm{AOA}}}} \right){{\bf{\Phi }}_m}[q],
		\end{array}
	\end{equation}
where $D_{\rm m}$ represents the maximum directivity of the STAR-RIS, $G_{k}$ signifies the receiving gain from UD $k$ to the STAR-RIS, and $G_{\rm{B}}$ denotes the transmitting gain from the STAR-RIS to BS. The matrix $\mathbf{\Phi}_m$ equals  $\mathbf{\Phi_{\rm r}}$ when UD $k$ in the RA and $\mathbf{\Phi_{\rm t}}$ when in the TA. Additionally, $F\left(\upsilon,\vartheta\right)$ indicates the normalized power radiation pattern of the STAR-RIS, with $\upsilon$ and $\vartheta$ denoting the azimuth and elevation angles between UD $k$ (BS) and the STAR-RIS, respectively. This can be modeled using an exponential-Lambertian radiation pattern parameterized by $z$, which is given by
	\begin{equation}
		F\left(\upsilon,\vartheta\right) = \cos^z\left( \upsilon \right) \cos^z\left( \vartheta \right),  \upsilon \in \left[ 0, 2\pi \right], \, \vartheta \in \left[ 0, \pi \right],
	\end{equation}
	Based on equations (7), (8), and (9), the STAR-RIS gain for UD $k$ can be rewritten as
	\begin{equation}
		\scalebox{0.99}{$
		{\bm{\xi} _k} = D_{\rm{m}}^2{\left| {\sin {\theta _k[q]}\cos {\varphi _k[q]}\sin {\theta _{\rm{B}}[q]}\cos {\varphi_{\rm{B}}[q]}} \right|^z}\mathbf{\Phi}_m[q].
		$}
	\end{equation}
	
Finally, we define ${h_k}[q] = {\boldsymbol{v}_{\rm{R},\rm{B}}}{[q]^{\rm H}}\boldsymbol{\xi}_k[q] {\boldsymbol{h}_{k,\rm{R}}}[q] + {h_{k,\rm{B}}}[q]$. By applying the orthogonal frequency division multiple access protocol for task offloading, the achievable rate for UD $k$ in time slot $q$ is given by
	\begin{equation}
		{R_k}[q] = {B_k}{\log _2}\left( {1 + \frac{{{p_k}[q]{{\left| {{h_k}[q]} \right|}^2}}}{{{\sigma ^2}}}} \right),
	\end{equation}
where $\sigma ^2$ signifies the Gaussian noise power, and ${p_k}[q]$ is the transmission power of UD $k$, ${B_k=B/K}$ denotes the bandwidth allocated to UD $k$, with $B$ signifying the total available bandwidth.

	\subsection{Computation Model}
	At the beginning of cycle $T$, UD $k$ is assigned a task ${M_k} = \{ {D_k},{C_k}\}$, where ${D_k}$ represents the task size and ${C_k}$ denotes the required CPU cycles. UDs perform partial offloading, splitting tasks between local and edge computing. Since the BS is equipped with high-performance servers, its energy consumption is considered negligible. Therefore, the system's energy consumption is mainly concentrated on task offloading and local computing by UDs.
	
	 \subsubsection{Task offloading}Since task offloading and execution require time, we consider that the MEC server does not execute tasks in the first time slot, and UDs do not offload tasks during the $Q$-th time slot. UD $k$ offloads a portion ${\alpha _k}[q] \in [0,1]$ of task size ${D_k}$, satisfying 
	 \begin{equation}
	 	\eta_k=\textstyle \sum\nolimits_{q = 1}^Q {{\alpha _k}[q]}\le 1,
	\end{equation}
	 so that $(1 - \eta_k){D_k}$ will be processed locally. The time and energy consumption for task offloading for UD $k$ are described as
	\begin{equation}
		t_k^{\rm{off}}[q] = {\alpha _k}[q]{D_k}{({R_k}[q])^{ - 1}},\forall q \in\mathcal Q,
	\end{equation}
	\begin{equation}
		E_k^{{\rm{off}}}[q] = t_k^{{\rm{off}}}[q]{p_k}[q],\forall q \in \mathcal Q.
	\end{equation}
	
	Once the offloaded tasks are received, the MEC server executes them in parallel. Let $f_k^{\rm{e}}$ represent the computation resource allocated to UD $k$ during cycle $T$, subject to
	\begin{equation}
	 	\sum\nolimits_{k = 1}^K {f_k^{\rm{e}}}  \le f_{{\rm{total}}}^{\rm{e}},
	\end{equation}
	where $f_{{\rm{total}}}^{\rm{e}}$ denotes the total computation resource of the MEC server. Consequently, the computation time for offloaded tasks from UD $k$ is determined by
	\begin{equation}
		t_k^{\rm{e}} = {D_k}{C_k}{\eta_k}/f_k^{\rm{e}}.
	\end{equation}
	\subsubsection{Local processing}The local computation time and energy consumption of UD $k$ during cycle $T$ are given by
	\begin{equation}
		t_k^{{\rm{loc}}} = {D_k}{C_k}(1 - {\eta_k})/f_k^{{\rm{loc}}},
	\end{equation}	
	\begin{equation}
		E_k^{{\rm{loc}}} = {c_k^{{\rm{loc}}}}{\left( {f_k^{{\rm{loc}}}} \right)^2}{D_k}{C_k}(1 - {\eta_k}),
	\end{equation}
	where $f_k^{{\rm{loc}}}$ and $c_k^{{\rm{loc}}}$ denote the CPU frequency and effective capacitance coefficient of UD $k$, respectively. Therefore, the energy consumption of UD $k$ is expressed as
	\begin{equation}
		{E_k} = E_k^{{\rm{loc}}} + \sum\nolimits_{q = 1}^{Q - 1} {E_k^{{\rm{off}}}[q]}.
	\end{equation}
	\subsection{Problem Formulation}
	Our objective is to minimize the total energy consumption through the joint optimization of STAR-RIS configurations, orientation, computation resource allocation, transmission power, and offloading strategies, while satisfying the constraints on latency and computation resource. Based on the characteristics of the STAR-RIS, this problem is formulated for all three operating protocols.
	
	\subsubsection{Problem formulated for ES/MS protocol}
	The energy consumption minimization problem for the ES/MS protocol can be formulated as
	\begin{subequations}
		\begin{align}
			&\quad\:\mathop {\min }\limits_{\mathbf z} \;\sum\limits_{k = 1}^K {{E_k}} \\
			&{\rm{s}}{\rm{.t}}{\rm{.}}\,\beta_n^{\rm{r}}[q],\beta _n^{\rm{t}}[q] \in [0,1],\forall n \in {\cal N},\forall q \in {\cal Q},\\
			&\quad\text\;\:\beta^{\rm t}_n[q]+\beta^{\rm r}_n[q]=1, \forall n \in {\cal N},\forall q \in \mathcal Q,\\
			&\quad\;\:\beta_n^{\rm{r}}[q],\beta _n^{\rm{t}}[q] \in \{0,1\},\forall n \in {\cal N},\forall q \in {\cal Q} \text{ (for MS),}\\
			&\quad\;\:f_k^{{\rm{loc}}} \in [0,f_{\max }^{{\rm{loc}}}],\forall k \in \mathcal K, \\
			&\quad\;\;{p_k}[q] \in [0,{p_{\max }}],\forall k \in \mathcal K,\forall q \in \mathcal Q, \\
			&\quad\;\:\max \{ t_k^{{\rm{loc}}},t_k^{{\rm{off}}},t_k^{\rm{e}}\}  \le T,\forall k \in \mathcal K, \\
			&\quad\;\; (3),(10),(15),(18),
		\end{align}
	\end{subequations}
where $\mathbf{z} = \{ \delta ,{{\bf{\Phi }}_{\rm{t}}},{{\bf{\Phi }}_{\rm{r}}},{\boldsymbol \alpha},{\boldsymbol p},\boldsymbol f^{{\rm{loc}}},\boldsymbol f^{\rm{e}}\}$ represents the set of optimization variables, with ${\boldsymbol \alpha}  = \left\{ {{\alpha _1}, \ldots ,{\alpha _K}} \right\}$, ${\boldsymbol p} = \left\{ {{p_{1,}} \ldots ,{p_K}} \right\}$, ${\boldsymbol f^{{\rm{loc}}}} = \left\{ {f_1^{\rm loc}, \ldots ,f_K^{\rm loc}} \right\}$, ${{\boldsymbol f}^{\rm{e}}} = \left\{ {f_1^{\rm e}, \ldots ,f_K^{\rm e}} \right\}$. Constraints (3), (10), (23e), and (23f) define the ranges for rotation angle, discrete phase shift, local computation resource, and transmission power. Constraints (31b) and (23c) jointly regulate the amplitude coefficients for STAR-RIS under the ES protocol. Constraints (23c) and (23d) govern the operating mode for each elements under the MS protocol. Constraint (23g) guarantees that tasks are completed within cycle $T$, where $t_k^{{\rm{off}}} = \sum\nolimits_{q = 1}^{Q-1} {t_k^{{\rm{off}}}[q]}$ denotes the sum offloading time across all time slots. Constraint (15) prevents the offloaded tasks from exceeding the total task size. Constraint (18) ensures that the allocated computation resource does not exceed the total available resource.
	
	\subsubsection{Problem formulated for TS protocol} The energy minimization problem for the TS protocol is formulated as 
	\begin{subequations}
		\begin{align}
			&\mathop {\min }\limits_{{\mathbf z}} \;\sum\limits_{k = 1}^K {{E_k}} \\
			&\quad{\rm{s}}{\rm{.t}}{\rm{.}}\,\lambda_{\rm t}[q],\lambda_{\rm r}[q] \in \{ 0,1\} , \forall q \in \mathcal Q,\\
			&\quad\quad\text\;\:\lambda_{\rm t}[q]+\lambda_{\rm r}[q]=1,\forall q \in \mathcal Q,\\
			&\quad\quad\;\; \rm{(23e)}\text{-}\rm{(23h).}
		\end{align}
	\end{subequations}
	Constraints (24b) and (24c) jointly ensure that the STAR-RIS is operated in T or R mode in each time slot.

	The main challenge in solving problems (23) and (24) stems from the strong coupling between variables, such as the STAR-RIS configurations and orientation, as well as the non-convexity of the optimization objective and constraints. Additionally, the mobility of UDs introduces further uncertainty into the optimization process. These factors make problems difficult to solve using traditional convex-based methods. Therefore, we adopt an SAC-based algorithm to tackle these challenges, which will be detailed in the next section.
	
	\section{Problem Solution}\label{pro:s}
	In this section, we first model the formulated energy consumption minimization problem by a Markov decision process (MDP) framework, and then introduce a sample-efficient SAC algorithm to solve it.
	\subsection{MDP Formulation}
	In implementing DRL, we begin by defining MDP which serves as the core structure for addressing sequential decision-making in a stochastic environment. An MDP is described by a five-tuple $\left\{ {\mathcal{S},\mathcal{A},\mathcal{P},\mathcal{R},\gamma} \right\}$, 
	where $\mathcal{S}$ is the set of environment states, $\mathcal{A}$ denotes the set of actions, $\mathcal{P}$ signifies the state transition probability matrix, $\mathcal{R}$ represents the reward function, and $\gamma$ indicates the discount factor. At each time slot, the agent selects an action $a \in \cal{A}$ from state $s \in \cal{S}$ according to its policy $\pi$. Specifically, the policy gives the probability distribution over actions for the given state, represented as $\pi (a|s) = P\left[ {{A_q} = a|{S_q} = s} \right] \in \left[ {0,1} \right]$. After executing action $a$, the agent transitions to the next state and receives a reward $r$. The specific definitions of the state, action, reward, and state transition within the formulated MDP are provided below.
	\subsubsection{State}
	The environment state at time slot $q$ is represented as ${s_q} = \left\{ {\boldsymbol\theta [q],\boldsymbol\varphi [q], \boldsymbol\alpha^{\rm{cu}} [q]}\right\}$, which is composed of three part:
	\begin{itemize}
		\item \textbf{} $\boldsymbol {\theta}[q]  = \{ {\theta _k}[q],k \in \mathcal K\}$ is the angles between UDs and STAR-RIS projected onto XOY plane at time slot $q$; 
		\item \textbf{} $\boldsymbol {\varphi}[q]  = \{ {\varphi_k}[q],k \in \mathcal K\}$ is the elevation angles between UDs and STAR-RIS at time slot $q$;
		\item \textbf{} ${\boldsymbol{\alpha} ^{{\rm{cu}}}}[q] = \left\{ {\sum\nolimits_{i = 1}^{q - 1} {{\alpha _k}[i]} ,k \in {\cal K}} \right\}$ is the cumulative ratios of offloaded tasks from the beginning to time slot $q-1$.
	\end{itemize}
	
	\subsubsection{Action}
	The action space in the formulated MDP involves STAR-RIS configurations, orientation, resource allocation, transmission power, and task offloading strategies at each time slot. Finding the optimal policy to govern all these variables presents significant challenges for several reasons. Firstly, practical constraints dictate that each STAR-RIS element can only offer a finite set of discrete phase shift and must switch between T or R mode in TS/MS mode, making a hybrid action space consisting of both discrete and continuous components. Secondly, the extensive action space and environmental uncertainties significantly complicate solving the MDP. To address these challenges, we first project the discrete variables into continuous spaces, converting the overall action space into a continuous domain. Moreover, to keep the action space manageable, we propose a solution for computing  transmission power and resource allocation based on given STAR-RIS configurations, orientation, and offloading strategies, as detailed below.
	
	According to equation (15), the local computation energy $E_k^{\rm {loc}}$ increases with $f_k^{\rm {loc}}$. Thus, $E_k^{\rm {loc}}$ is minimized when $f_k^{\rm {loc}}$ reaches its minimum value. From equation (14), when local computation time $t_k^{\rm {loc}}=T$, $f_k^{\rm {loc}}$ achieves its optimal value, which can be expressed as
	\begin{equation}
		{ {f_k^{{\rm{loc}}}} ^*} = {D_k}{C_k}\left( {1 - {\eta _k}} \right)/T.
	\end{equation}
	
	Similarly, to execute the offloaded tasks within $Q-1$ time slots, the optimal computation resource allocation is given by
	\begin{equation}
		{ {f_k^{\rm{e}}}^*} = {D_k}{C_k}{\eta _k}/\left( {\left( {Q - 1} \right)\tau } \right).
	\end{equation}
	
	Based on equation (13), we can demonstrate that $t_k^{\rm {off}}[q]$ is inversely related to $p_k[q]$. Since the offloading time must satisfy $t_k^{{\rm{off}}}[q] \in [0,\tau]$, the transmission power must satisfy the following constraint
	\begin{equation}
		p_k[q] \ge {\sigma ^2}\left( {{2^{{\alpha _k}[q]{D_k}{{\left( {\tau {B_k}} \right)}^{ - 1}}}} - 1} \right){\left| {{h_k}[q]} \right|^{ - 2}} = \hat p_k[q].
	\end{equation}
	If ${{\hat p}_k}[q] > p_{\rm{max}}$, we set $p_k[q]=p_{\rm {max}}$, and calculate the maximum offloading ratio for UD $k$ in this time slot based on the current information. Otherwise, we combine equations (16), (17), and (27), then the problem of minimizing transmission energy consumption through optimizing transmission power is formulated as
	\begin{subequations}
		\begin{align}
			&\quad\text{ }\mathop {\min }\limits_{{p_k[q]}} \;E_k^{{\rm{off}}}[q] = \frac{{{\alpha _k}[q]{D_k}[q]{p_k}[q]}}{{B_{k}{{\log }_2}\left( {1 + {p_k}[q]{{\left| {{h_k}[q]} \right|}^2}{\sigma ^{-2}}} \right)}}\\
			&{\rm{s}}{\rm{.t}}{\rm{.}}\;{\hat p_k[q] \le p_k[q] \le {p_{\max }}}.
		\end{align}
	\end{subequations}
	
	The optimization problem concerning $p_k$ is a typical convex-concave ratio maximization problem, which can be efficiently solved using Dinkelbach's algorithm \cite{FP}. The objective can be reformulated (23a) as
	\begin{equation}
		\scalebox{0.92}{$
			\mathop {\min }\limits_{{p_k}[q]} \;{\alpha _k}[q]{D_k}[q]{p_k}[q] - {y_k}{B_k}{\log _2}\left( {1 + {p_k}[q]{{\left| {{h_k}[q]} \right|}^2}{\sigma ^{-2}}} \right),
			$}
	\end{equation}
	with a new auxiliary variable $y_k$, iteratively updated by
	\begin{equation}
		{{y_k}^{t + 1}}{\rm{ = }}\frac{{{\alpha _k}[q]{D_k}[q]{p_k^t}{{[q]}}}}{{B_{k}{{\log }_2}\left( {1 + {p_k^t}{{[q]}}{{\left| {{h_k}[q]} \right|}^2}{\sigma ^{-2}}} \right)}},
	\end{equation}
	where $t$ is the iteration index. Convergence is ensured by alternatively updating $y$ according to (25) and solving for $p_k[q]$ in (24), as $y$ is non-increasing with each iteration. The optimal transmitting power ${{{p_k}[q]}^*}$ can be obtained after successive iterations. The process is summarized in Algorithm 1.
	
	\begin{algorithm}
		\caption{Dinkelbach-Based Algorithm}
		\begin{algorithmic}[1]
			\STATE {Initialize:} $t = 1$, initial $p_k[q] = \hat{p}_k[q]$, $y_k^{t}$, and $\epsilon$
			\FOR{$t \le {t_{\max }}$}
			\STATE Optimize $p_k^t[q]$ by solving problem (29) with $y^t$;
			\STATE Calculate $y^{t+1}$ according to (30);
			\STATE \textbf{if} \scalebox{0.96}{$\left| {{\alpha _k}[q]{D_k}[q]p_k^t[q] - {y^t}{B_k}{{\log }_2}\left( {1 + \frac{{p_k^t[q]|{h_k}[q]{|^2}}}{{{\sigma ^2}}}} \right)} \right| \le \epsilon$}
			\STATE \hspace{1em} \textbf{break};
			\ENDFOR
		\end{algorithmic}
	\end{algorithm}

By utilizing equations (25) and (26), along with solving problem (28), the optimal transmission power and computation resource allocation can be determined for given STAR-RIS configurations, orientation, and task offloading strategies. Therefore, in the proposed MDP model, only the decisions for STAR-RIS configurations, orientation, and task offloading strategies need to be included in action space, while the optimal transmission power and resource allocation can be derived for state-value calculation. However, for different operating protocols, the action space is also variable, which is detailed as below.
	
	\textbf{Action space for ES/MS protocol:} The action space for ES/MS protocol in each time slot includes following five parts:
	\begin{itemize}
		\item The rotation angle of the STAR-RIS, i.e. $\delta[q]$;
		\item The reflection phase shift coefficient of each sub-surface, i.e. $\phi_{\bar n}^{\rm r}[q]$;
		\item The transmission phase shift coefficients of each sub-surface, i.e. $\phi_{\bar n}^{\rm t}[q]$;
		\item The reflection amplitude coefficients of each sub-surface, i.e. $\beta_{\bar n}^{\rm r}[q]$;
		\item The offloading ratio of each UD, i.e. $\alpha_k[q]$.
	\end{itemize}
	
Considering the discrete phase shift, the action space combines both discrete and continuous components, which complicates the solution of the formulated MDP. To address this, we divide the continuous space $[0,2^{b}-1]\pi$ into $2^b$ smaller segments, each with length $\varsigma  = \frac{{{\pi}}}{{{2^{b-1}}}}$. Furthermore, under the MS protocol, each element can only be operated in either T or R mode, resulting in a binary limitation on reflection and transmission amplitude coefficients. A similar approach is applied to handle this binary constraint. Consequently, the size of action space for ES/MS protocol is $3\bar N+K+1$.

\textbf{Action space for TS protocol:}
Considering that under the TS protocol the STAR-RIS can only be operated in either T or R mode during a given time slot, only UDs in one half-space are served, and one phase shift matrix is used. Defining the action space similar to the ES/MS protocol would cause inefficiency, resulting in slower convergence. Notably, UD $k$ is located in the RA when $u_k[q]=1$, and when $\lambda_r[q]=1$ the STAR-RIS is operated in reflection mode. Therefore, whether UD $k$ is served in time slot $q$ can be determined by
	\begin{equation}
		{i_k}[q] = \lambda_{\rm r}[q]\bar\oplus {u_k}[q],
	\end{equation}
	where $\bar\oplus$ denotes the exclusive NOR operation. The STAR-RIS gain for UD $k$ can be reformulated as
	\begin{equation}
		\scalebox{1}{$
			{\bm{\xi} _k} = {i_k[q]}D_{\rm{m}}^2{\left| {\sin {\theta _k[q]}\cos {\varphi _k[q]}\sin {\theta _{\rm{B}}[q]}\cos {\varphi_{\rm{B}}[q]}} \right|^z}\mathbf{\Phi}[q],
			$}
	\end{equation}
	where $\mathbf{\Phi}$ is the phase shift matrix corresponding to the STAR-RIS operating mode. Consequently, the action space involves following four parts:
	\begin{itemize}
		\item The rotation angle of the STAR-RIS, i.e. $\delta[q]$;
		\item The phase shift coefficients for transmission/reflection of each sub-surface, i.e. $\phi_{\bar n}[q]$;
		\item The operating mode, i.e. $\lambda_{\rm r}[q]$;
		\item The offloading ratio of each UD, i.e. $\alpha_k[q]$.
	\end{itemize}
	The size of action space under TS protocol is $\bar N+K+2$.
	\subsubsection{Reward}
	The goal is to minimize the total energy consumption, while the agent aims to maximize its reward. To ensure the reward aligns with this goal, the instantaneous total energy consumption for task offloading of all UDs at each time slot is incorporated into the reward, namely $\bar{E}[q] = \sum\nolimits_{k = 1}^K {E_k^{{\rm{off}}}} [q]$. In response to MEC computation resource constraint, a penalty $P_1$ is applied if the allocated resource exceeds the maximum. For the time constraint, a penalty $P_2$ is given if tasks are not completed by the end of cycle $T$. Then, the reward function is defined as
	\begin{equation}
		\scalebox{0.95}{$
		\arraycolsep=1.4pt 
		r(s_q, a_q) = \left\{
		\begin{array}{ll}
			- \bar{E}[q], & q \leq Q - 1, \\
			- \sum\limits_{k = 1}^K E_k^{\text{loc}} - \bar{E}[q] - P_1-P_2, & q = Q - 1,
		\end{array}
	\right.
	$}
	\end{equation}
	where $P_1 = \left( {\sum\nolimits_{k = 1}^K {f_k^{\rm{e}}}  - f_{{\rm{total}}}^{\rm{e}}} \right)W$ and $P_2=W$, with $W$ denoting a positive constant.

	\subsubsection{State Transition}
	The cumulative ratios of offloaded tasks are updated by
	\begin{equation}
		\alpha _k^{{\rm{cu}}}[q] = \alpha _k^{{\rm{cu}}}[q - 1] + {\alpha _k}[q - 1].
	\end{equation}
	The angles between UDs and STAR-RIS, both projected onto the XOY plane and in elevation, are calculated by equations (1) and (6), respectively.
	\subsection{SAC-Based Algorithm}
	The long-anticipated application of DRL frameworks has progressed slowly in practice, mainly because of low sampling efficiency and fragile convergence \cite{ref22}. To address these challenges, SAC has been proposed, leveraging a maximum entropy framework to enhance sampling efficiency. The SAC algorithm consists of three main components: the rotatable STAR-RIS-assisted MEC environment, the SAC agent, and a replay buffer, as depicted in Fig. 4. Based on the current observed state $s_t$, the SAC agent selects the corresponding action $a_t$ \cite{ref21}. The environment then calculates the immediate reward $r_t$ according to $a_t$, and transitions to the next state $s_{t+1}$. The experience tuple $\{ {s_t},{a_t},{r_t},{s_{t + 1}}\}$ is saved in the replay buffer. During training, the SAC agent retrieves a mini-batch of experiences from the replay buffer for learning. Comparing to conventional DRL algorithms, SAC offers several advantages, such as enhancing exploration efficiency, the ability to find near-optimal policy in multiple modes, and accelerating learning speed, particularly in challenging scenarios. 
	
	\begin{figure}[t]
		\centering
		\includegraphics[width=\columnwidth]{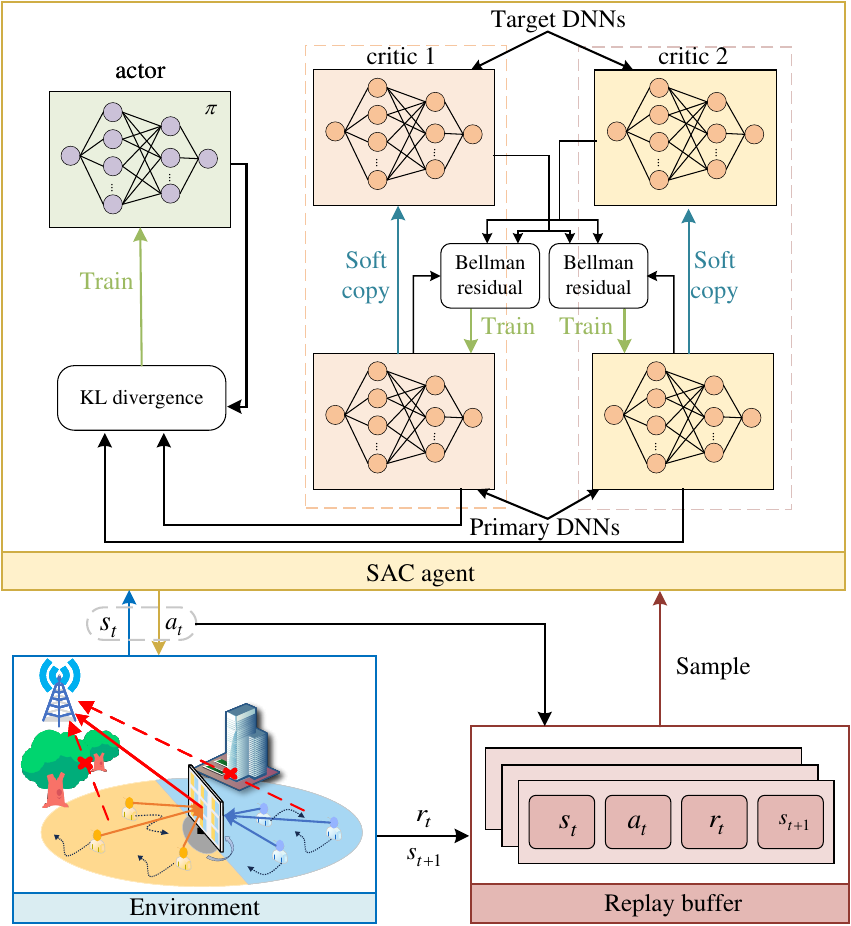}
		\captionsetup{justification=raggedright, singlelinecheck=false} 
		\caption{Architecture of SAC-based algorithm.}
		\label{fig:network}
	\end{figure}
	
	The goal of conventional DRL algorithms is to maximize the long-term return beginning from the initial state, given by
	\begin{equation}
		\mathop{\max}\limits_\pi \; \sum_{q=1}^{Q-1} \mathbb{E}_{(s_q, a_q) \sim \rho_\pi} \left[ \gamma^{q - 1} r(s_q, a_q) \right]
	\end{equation}
	where $\rho_\pi$ denotes the distribution of state-action trajectories under policy $\pi$. However, SAC incorporates an entropy term into the objective function to encourage exploration, defined as
	\begin{equation}
		\sum\limits_{q=1}^{Q-1} \mathbb{E}_{(s_q, a_q) \sim \rho_\pi} \left[ \gamma^{q - 1} r(s_q, a_q) + \alpha \mathcal{H}(\pi(\cdot| s_q)) \right],
	\end{equation}
	where $\mathcal H(\pi(\cdot | s_q)) = - \mathbb{E}_{a \sim \pi(\cdot | s_q)} \log_2 \pi(a | s_q)$ denotes the policy distribution entropy, while the temperature parameter $\alpha$ regulates the importance of entropy, reflecting the degree of stochasticity in the optimal policy ${{\rm{\pi }}^*}$. Since a fixed temperature may lead to instability due to the dynamic reward, the temperature $\alpha$ is necessary to be adjusted automatically. According to \cite{ref21}, the optimal dual variable $\alpha^*_q$ is calculated by
	\begin{equation}
		\scalebox{0.95}{$
			\alpha_q^* = \mathop{\arg\min}_{\alpha_q} \mathbb{E}_{a_q \sim \pi_q^*} \left[ -\alpha_q \log_2 (\pi_q^* (a_q \mid s_q);\alpha_q) - \alpha_q \mathcal{H}_{\min} \right],
			$}
	\end{equation}
	where ${\pi _q^ * \left( {{a_q}\left| {{s_q}} \right.} \right)}$ represents the optimal policy corresponding to $\alpha_q$, and $\mathcal H_{\rm min}$ represents the minimum entropy constraint.
	
	The SAC framework is fundamentally based on the policy iteration algorithm, comprising two primary phases: policy evaluation and improvement. In the evaluation phase, the action values for a given policy $\pi$ are assessed using the Bellman expectation function, expressed as $Q_\pi(s_q, a_q) = r(s_q, a_q) + \gamma \mathbb E_{s_{q+1} \sim p_s} \left[ v_\pi(s_{q+1}) \right]$. Unlike conventional DRL algorithms, by incorporating the entropy, the state-value function of SAC is given by
	\begin{equation}
		{v_\pi }\left( {{s_q}} \right) = {\mathbb E_{{a_q} \sim \pi }}\left[ {{Q_\pi }\left( {{s_q},{a_q}} \right) - \alpha {{\log }_2}\left( {\pi \left( {{a_q}\left| {{s_q}} \right.} \right)} \right)} \right].
	\end{equation}
	
	Given the continuous state space in the proposed MDP model, deep neural networks (DNNs) are utilized to approximate state values. With the Q-network parameterized by $\omega$, the loss function is formulated as
	\begin{equation}
		\scalebox{1}{$
		{L_Q}\left( \omega  \right) = {\mathbb E_{\left( {{s_q},{a_q}} \right) \sim {\cal D}}}\left[ {\frac{1}{2}{{\left( {{Q_\omega }\left( {{s_q},{a_q}} \right) - {Q_{\hat \omega }}\left( {{s_q},{a_q}} \right)} \right)}^2}} \right],
	$}
	\end{equation}
	where $\cal D$ represents the reply buffer, while $\omega $ and ${\hat \omega }$ denote the parameters of the Q-network and target Q-network, respectively. Similarly, the policy network loss function is defined as  
	\begin{equation}
		\scalebox{1}{$
		{L_\pi }\left( \varphi  \right) = {\mathbb{E}_{{s_q} \sim {\cal D}}}{\mathbb{E}_{{a_q} \sim {\pi_\varphi}}}\left[ \alpha {{\log}_2}\!\left( {\pi_\varphi}\left( {a_q\left| {s_q} \right.} \right) \right) - {Q_\omega }\left( {s_q},{a_q} \right) \right],
		$}
	\end{equation}
	where $\varphi $ denotes the parameter of the policy-network. To further accelerate training and enhance stability, this paper integrates prioritized experience replay into the SAC algorithm. Fig. 4 and Algorithm 2 illustrate the architecture and training process of proposed SAC-based algorithm.
	
	\begin{algorithm}[t]
		\caption{Energy Consumption Minimization Algorithm Based on SAC}
		\label{SAC}
		\begin{algorithmic}[1]
			\STATE{Initialize environment;}
			\STATE{
				Initialize ${\omega _i}(i = 1,2)$ for critic networks, $\varphi$ for actor network;}
			\STATE{
				Initialize entropy level $\mathcal H_{\min }$, replay buffer $D = \emptyset$, learning rate and temperature parameter respectively;}
			\FOR{each epoch}
			\FOR{each step}
			\STATE{Take action based on current policy $\pi_{\varphi}$;}
			\STATE{Calculate reward $r\left( s_q,a_q \right)$, $f_k^{\rm loc}$, $f_k^{\rm e}$ by equations (33), (25), and (26) respectively, obtain $p_k$ by solving problem (29), and observe next state $s_{q+1}$;}
			\STATE{Store transition $\left\{ s_q, a_q, r(s_q,a_q), s_{q+1} \right\}$ in $\cal D$, and update UDs' location;}
			\ENDFOR
			\FOR{each gradient step}
			\STATE{Prioritized sample a mini-batch of experience from $\cal D$;}
			\STATE{Update critic networks $\omega_i$ by loss function (34);}
			\STATE{Update the actor network $\varphi$ by loss function (35);}
			\STATE{Update temperature $\alpha$ by solving (37);}
			\STATE{Update target network parameter $\hat \omega_i$ periodically;}
			\ENDFOR
			\ENDFOR
			\RETURN{The optimal policy $\rm{\pi}_{\varphi}^{*}$}
		\end{algorithmic}
	\end{algorithm}
	
	\subsection{Complexity Analysis}
In this part, we evaluate the computation complexity of the proposed algorithm, focusing on training actor and critic networks, as well as action generation. The state space dimension equals $3K$. The first and second hidden layers have $L_1$ and $L_2$ neurons, respectively. The action space dimensions for each operating protocol are $3\bar N+K+1$ (for ES/MS), and $\bar N+K+1$ (for TS). To simplify analysis, let the size of action space be denoted by $A$. According to [36], the complexity for primary DNNs of the actor and each critic equals $\mathcal{O}\left( {{L_1} + {L_1}{L_2} + {L_2}\left( A \right)} \right)$, and $\mathcal{O}\left( {{L_1}\left( {A + 3K} \right) + {L_1}{L_2} + {L_2}} \right)$, respectively. The critic's target DNNs are updated using soft copy, making their complexity negligible. Therefore, the algorithm complexity for training can be represented as $\mathcal{O}\left( {\Gamma {\rm{len}}\left( {2\left( {{L_1} + \frac{{{L_2}}}{2}} \right)A + 3{L_1}\left( {{L_2} + 3K} \right) + 2{L_2}} \right)} \right)$, where $\Gamma$ denotes the maximum training episodes, len is the episode length. After training, the actor can generate actions directly, with a generation complexity of $\mathcal{O}\left( {{L_1} + {L_1}{L_2} + {L_2}A} \right)$.

	\section{Simulation Results}
	In this section, numerical simulations are conducted to validate the effectiveness and superiority of the proposed algorithm in the rotatable STAR-RIS-assisted MEC system. For comparison, the following benchmarks are used:
	
	\begin{itemize}
		\item \textbf{Proximal Policy Optimization (PPO):} This method is a popular and reliable DRL algorithm that uses a stochastic policy, which defines a distribution over actions instead of providing a deterministic policy. PPO employs a clipped objective function to ensure stable updates, preventing drastic change to the policy and enhancing training stability \cite{PPO}.
		\item \textbf{Deep Deterministic Policy Gradient (DDPG):} This algorithm combines deep learning with deterministic policy methods, designed for scenarios involving high-dimensional state and continuous action spaces. DDPG also employs experience replay for better sample efficiency and stability during training, making it effective in high-dimensional environments \cite{DDPG}.
		\item \textbf{Fixed STAR-RIS:} This scheme is designed for demonstrating the performance improvement achieved through rotation by keeping STAR-RIS orientation constant, where the STAR-RIS is oriented to the BS. By maintaining this alignment, we can effectively evaluate the potential of orientation optimization in enhancing MEC performance.
		\item \textbf{Transmitting/Reflecting-Only RIS:} To demonstrate the advantages of STAR-RIS, this scheme employs transmitting-only and reflecting-only RIS to assist UDs in task offloading.
		\item \textbf{Local Process:} UDs execute tasks locally without offloading, with $\alpha_k[q]=0$.
	\end{itemize}
	
	\begin{table}[htbp]
		\renewcommand{\arraystretch}{1.2} 
		\centering
		\caption{\scriptsize {EXPERIMENTAL PARAMETERS}}
		\begin{tabular}{|>{\centering\arraybackslash}p{0.32\textwidth}|>{\centering\arraybackslash}p{0.12\textwidth}|}
			\hline
			Parameters & Values \\ \hline
			$\text{The maximum transimitting power }{p_{\max }}$ & $0.2 \text{ }\mathrm{W}$ \\ \hline
			$\text{The number of UDs } K$ & $6$\\ \hline
			$\text{The task cycle } T$ & $10 \text{ }\mathrm{s}$\\ \hline
			$\text{Time slots } Q$ & $5$\\ \hline
			$\text{The bandwidth }B$ & $5 \text{ }\mathrm{MHz}$ \\ \hline
			$\text{The channel Gaussian white Noise }{\sigma ^2}$ & $-110 \text{ }\mathrm{dBm}$ \\ \hline
			$\text{Path loss exponent }{\alpha_1, \alpha_2}$ & $2$ \\ \hline
			$\text{Racian factors }{K_1}$, ${K_2}$ & $10$ \\ \hline
			$\text{The phase quantization bit }b$ & $2\text{ }\mathrm{bit}$ \\ \hline
			$\text{Effective capacitance }c_k^{\mathrm{loc}}$ & $10^{-27}\text{ }\mathrm{J/cycle}$ \\ \hline
			$\text{The task size }{D_k}$ & $10 \text{ } \mathrm{Mb}$ \\ \hline
			$\text{The require computation resource per bit }{C_k}$ & $600 \text{ }\mathrm{cycles/bit}$ \\ \hline
			$\text{Total computation resource  }f_{\mathrm{total}}^{\mathrm{e}}$ & $10 \text{ }\mathrm{GHz/s}$ \\ \hline
			$\text{The maximum computation resource of UD }k f_{\max}^{\mathrm{loc}}$ & $0.6\text{ } \mathrm{GHz/s}$ \\ \hline
			$\text{The radiation pattern parameter } z$ & $2$\\ \hline
			$\text{The movement velocity of UDs  } v_k$ & $[1.1,1.5]\text{ }\mathrm{m/s}$\\ \hline
			$\text{The positive constant }W$ & $10$ \\ \hline
		\end{tabular}
		\label{tab:parameters}
	\end{table}

		\begin{table}[htbp]
			\renewcommand{\arraystretch}{1.2} 
			\centering
			\caption{\scriptsize {HYPERPARAMETERS OF THE ALGORITHM}}
			\begin{tabular}{|>{\centering\arraybackslash}p{0.32\textwidth}|>{\centering\arraybackslash}p{0.12\textwidth}|}
				\hline
				Parameters & Values \\ \hline
				${\text{The episode length len}}$ & $100$ \\ \hline
				The maximum training episodes $\Gamma$ & $5000$ \\ \hline
				The replay pool size & $10^6$ \\ \hline
				The actor learning rate  & $1 \times 10^{-4}$ \\ \hline
				$\text{The discount factor }\gamma$ & $0.99$ \\ \hline
				The critic learning rate  & $1 \times 10^{-4}$ \\ \hline
				The number of samples of per batch & $256$ \\ \hline
				Replay memory size & $10^6$ \\ \hline
				Importance sampling exponent & $0.6$ \\ \hline
				$\text{Optimizer }$ & Adam \\ \hline
			\end{tabular}
			\label{tab:hyperparameters}
		\end{table}
		
	\subsection{Simulation Settings}
	 Here are the details of the simulation setting. UDs are randomly distributed within a concentric circular region with radii ranging from 2 to 7 meters, centered at (50, 0, 0) meters. The BS and STAR-RIS are positioned at (0, 0, 10) and (50, 0, 1) meters, respectively. Unless otherwise state, the experimental parameters in our simulation and the hyperparamaters for the SAC-based algorithm are depicted in  TABLE III and TABLE IV,  respectively.

			\begin{figure}[t]
				\centering
				\begin{subfigure}[t]{0.48\textwidth}
					\centerline{\includegraphics[width=\textwidth]{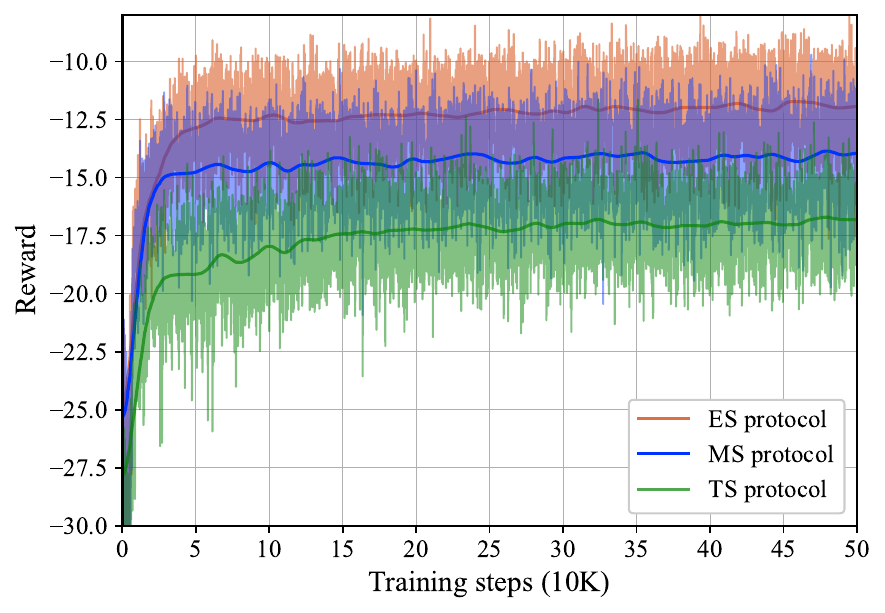}}
					\caption{Convergence performance for three protocols.}
					\label{fig:convergence}
				\end{subfigure}
			\end{figure}
			
	\subsection{Performance Evaluation}
	
	Fig. 5(a) shows the learning speed of the SAC-based algorithm under different operating protocols. It is evident that the proposed algorithm converges rapidly, stabilizing at about 100K steps. However, the algorithm shows relatively slower convergence under the TS protocol. This arises from the binary nature of the operating mode in the TS protocol, which, although converted to a continuous domain, still poses challenges for the policy network in representing the optimized action distribution, thereby slowing learning speed. Furthermore, to demonstrate the superiority of the proposed SAC algorithm, we compare it with PPO and DDPG under three protocols in Fig. 5. SAC consistently outperforms the other algorithms in terms of both convergence and performance across all protocols, indicating efficient learning process and strong adaptation to the dynamic environment. For instance, under the ES protocol, SAC stabilizes in approximately 50K steps, whereas PPO takes around 100K steps, and DDPG fails to converge by the end of training. Under the TS protocol, both PPO and DDPG exhibit poor performance and are almost unable to learn effective strategies. The superior performance of SAC can be attributed to the incorporation of entropy into its learning objectives, encouraging broader exploration and enhancing adaptability.

	\begin{figure}[t]
		\centering
		\begin{subfigure}[t]{0.48\columnwidth}
			\centerline{\includegraphics[width=\textwidth]{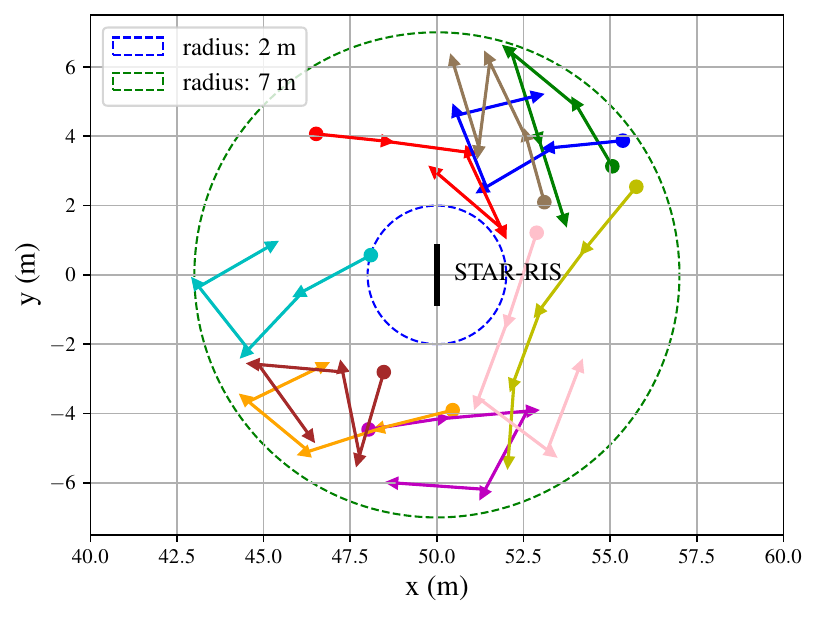}}
			\caption{UDs movement trajectory.}
			\label{fig:UDs movement}
		\end{subfigure}
		\begin{subfigure}[t]{0.48\columnwidth}
			\centerline{\includegraphics[width=\textwidth]{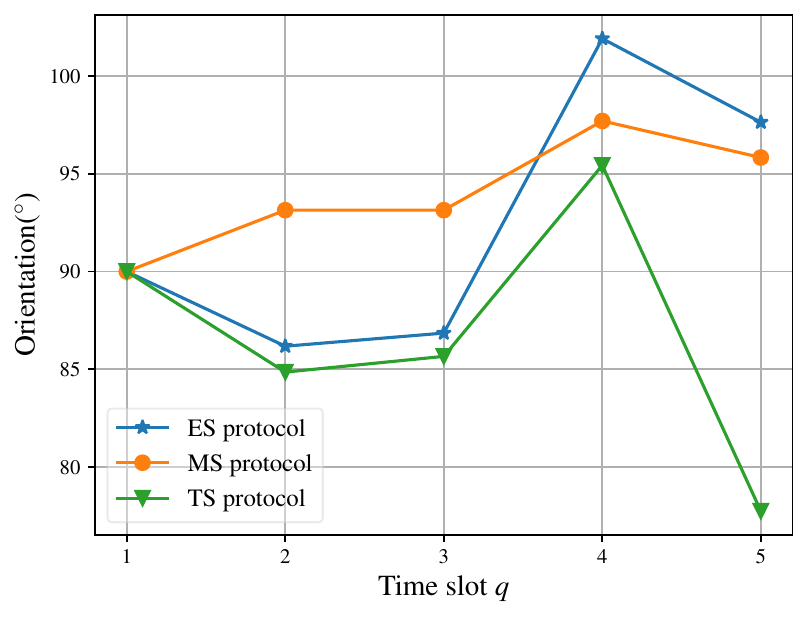}}
			\caption{Orientation variation.}
			\label{fig:Orientation}
		\end{subfigure}
		\caption{UDs movement trajectory and orientation variation of the proposed rotatable STAR-RIS under three operating protocol, with $K=10$ and seed=2024.}
	\end{figure}
	
	\begin{figure}[t]
		\centering
		\includegraphics[width=0.95\columnwidth]{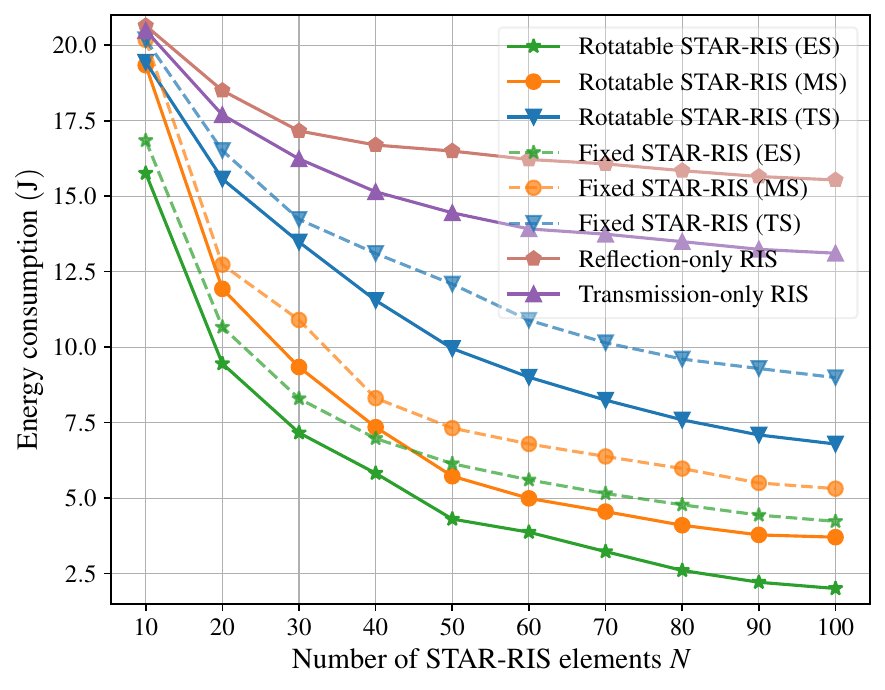}
		\caption{The energy consumption performance VS the number of STAR-RIS elements, with $K=10$ and seed=2024.}
		\label{fig:performance}
	\end{figure}

	\begin{figure}[htbp]
		\centering
		\includegraphics[width=0.95\columnwidth]{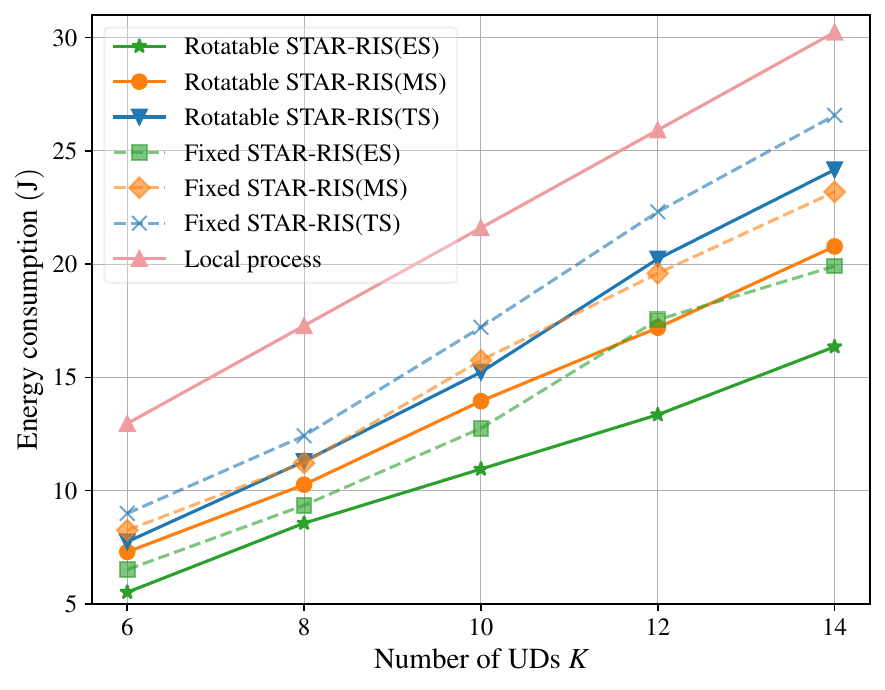}
		\caption{The energy consumption performance VS UDs number, with seed ranging from 2020 to 2024.}
		\label{fig:userNum}
	\end{figure}
	
	Fig. 6 depicts the movement trajectories of UDs and the orientation adjustment of the proposed STAR-RIS at each time slot under three protocols. In contrast to the fixed STAR-RIS, which is limited to adjusting its configurations to accommodate UDs mobility, the rotatable STAR-RIS can jointly optimize configurations and orientation, extending coverage to a broader range of UDs while maximizing STAR-RIS gain, as confirmed by the simulation results. Moreover, the rotation angles required by the STAR-RIS are minimal, while yielding significant performance improvement.
	
	Fig. 7 demonstrates that the energy consumption of UDs consistently decreases as the number of STAR-RIS elements $N$ increases. This trend is intuitive, as a large $N$ enhances the transmitting and receiving gains, facilitating higher offloading ratios, consequently reducing the energy consumption of UDs. The proposed rotatable STAR-RIS design outperforms the fixed STAR-RIS across all three operating protocols, and the gap between them widens progressively as $N$ increases. For example, under the ES protocol, the rotatable STAR-RIS reduces energy consumption by up to 52.7\% compared to the fixed STAR-RIS when $N=100$, which verifies the effectiveness of rotation. Additionally, with the increase of elements, the performance gain from rotation can even surpass the inherent limitations.

To verify the effectiveness of the proposed scheme for multiple moving UDs, we compare its performance against other schemes for varying numbers of UDs. Considering the randomness of UDs mobility, we consider random seeds ranging from 2020 to 2024, and calculate the average energy consumption. As shown in Fig. 8, the proposed rotatable STAR-RIS scheme outperforms the fixed STAR-RIS under all three protocols. This performance advantage is attributed to the ability of the rotatable STAR-RIS to dynamically adjust its orientation, optimizing both configuration and direction to accommodate the random movement of UDs. Moreover, as the number of UDs increases, the energy saving enabled by the rotatable STAR-RIS becomes even more significant, further demonstrating the scalability and efficiency of the proposed rotatable STAR-RIS in managing energy consumption for numerous UDs.

	\section{Conclusion}
	In this paper, we have investigated the rotatable STAR-RIS-assisted MEC systems with moving UDs under three operating protocols, focusing on the joint optimization of the STAR-RIS configurations, orientation, computation resource allocation, transmission power and task offloading strategies. We have proposed an SAC-based solution to deal with the optimization problem. Simulation results have demonstrated that the SAC-based approach outperforms other DRL algorithms in both convergence and overall performance. Moreover, even minor adjustments in the STAR-RIS orientation, based on the relative positions of UDs and BS, can yield significant performance gains, offering a promising direction for further research.

	\bibliographystyle{IEEEtran}
	\bibliography{references.bib}

\end{document}